\documentclass[conference]{IEEEtran}

\usepackage{amsmath}

\ifCLASSINFOpdf
  \usepackage[pdftex]{graphicx}
\else
\fi
\usepackage{array}

\usepackage{multirow}

\begin{document}
%
\title{SNRA: A Spintronic Neuromorphic Reconfigurable Array for In-Circuit Training and Evaluation of Deep Belief Networks}

\author{\IEEEauthorblockN{Ramtin Zand}
\IEEEauthorblockA{Department of Electrical and Computer Engineering\\
University of Central Florida\\
Orlando, FL 32816-2362\\
ramtinmz@knights.ucf.edu
}
\and
\IEEEauthorblockN{Ronald F. DeMara}
\IEEEauthorblockA{Department of Electrical and Computer Engineering\\
University of Central Florida\\
Orlando, FL 32816-2362\\
ronald.demara@ucf.edu}
}


%


\maketitle

\begin{abstract}
In this paper, a spintronic neuromorphic reconfigurable Array (SNRA) is developed to fuse together power-efficient probabilistic and in-field programmable deterministic computing during both training and evaluation phases of restricted Boltzmann machines (RBMs). First, probabilistic spin logic devices are used to develop an RBM realization which is adapted to construct deep belief networks (DBNs) having one to three hidden layers of size 10 to 800 neurons each. Second, we design a hardware implementation for the contrastive divergence (CD) algorithm using a four-state finite state machine capable of unsupervised training in N+3 clocks where N denotes the number of neurons in each RBM. The functionality of our proposed CD hardware implementation is validated using ModelSim simulations. We synthesize the developed Verilog HDL implementation of our proposed test/train control circuitry for various DBN topologies where the maximal RBM dimensions yield resource utilization ranging from 51 to 2,421 lookup tables (LUTs).  Next, we leverage spin Hall effect (SHE)-magnetic tunnel junction (MTJ) based non-volatile LUTs circuits as an alternative for static random access memory (SRAM)-based LUTs storing the deterministic logic configuration to form a reconfigurable fabric. Finally, we compare the performance of our proposed SNRA with SRAM-based configurable fabrics focusing on the area and power consumption induced by the LUTs used to implement both CD and evaluation modes. The results obtained indicate more than 80\% reduction in combined dynamic and static power dissipation, while achieving at least 50\% reduction in device count.
\end{abstract}


%
\IEEEpeerreviewmaketitle

\section{Introduction}

Within the post-Moore era ahead, several design factors and fabrication constraints increasingly emphasize the requirements for in-circuit adaptation to as-built variations. These include device scaling trends towards further reductions in feature sizes~\cite{Nikonov2015}, the narrow operational tolerances associated with the deployment of hybrid Complementary Metal Oxide Semiconductor (CMOS) and post-CMOS devices ~\cite{Ghosh2010,Ghosh2016}, and the noise sensitivity limits of analog-assisted neuromorphic computing paradigms \cite{Liu2013}. While many recent works have advanced new architectural approaches for the evaluation phase of neuromorphic computation utilizing emerging hardware devices, there have been comparatively fewer works to investigate the hardware-based realization of their training and adaptation phases that will also be required to cope with these conditions. Thus, this paper develops one of the first viable approaches to address post-fabrication adaptation and retraining in-situ of resistive weighted-arrays in hardware, which are ubiquitous in post-Moore neuromorphic approaches. Namley, a tractable in-field reconfiguration-based approach is developed to leverage in-field configurability to mitigate the impact of process variation. Reconfigurable fabrics are characterized by their fabric flexibility, which allows realization of logic elements at medium and fine granularities, as well as in-field adaptability, which can be leveraged to realize variation tolerance and fault resiliency as widely-demonstrated for CMOS-based approaches such as  \cite{Oreifej2018,Ashraf2013}. Utilizing reconfigurable computing by applying hardware and time redundancy to the digital circuits offers promising and robust techniques for addressing the above-mentioned reliability challenges. For instance, it is shown in \cite{Ashraf2013} that a successful refurbishment for a circuit with 1,252 look-up tables (LUTs) can be achieved with only 10\% spare resources to accommodate both soft and hard faults.   

Within the post-Moore era, reconfigurable fabrics can also be expected to continue their transition towards embracing the benefits of increased heterogeneity along several cooperating dimensions to facilitate neuromorphic computation~\cite{DeMara2017}. Since the inception of the first field-programmable devices, various granularities of general-purpose reconfigurable logic blocks and dedicated function-specific computational units have been added to their structures.  These have resulted in increased computational functionality compared to homogeneous architectures. In recent years, emerging technologies are proposed to be leveraged in reconfigurable fabrics to advance new transformative opportunities for exploiting technology-specific advantages. Technology heterogeneity recognizes the cooperating advantages of CMOS devices for their rapid switching capabilities, while simultaneously embracing emerging devices for their non-volatility, near-zero standby power, high integration density, and radiation-hardness. For instance, spintronic-based LUTs are proposed in \cite{Zandphys,Suzuki2015,Yang2018} as the primary building blocks in reconfigurable fabrics realizing significant area and energy consumption savings. In this paper, we extend the transition toward heterogeneity along various logic paradigms by proposing a heterogeneous technology fabric realizing both probabilistic and deterministic computational models. The cooperating advantages of each are leveraged to address the deficiencies of the others during the neuromorphic training and evaluation phases, respectively.  

In this paper, we propose a spintronic neuromorphic reconfigurable Array (SNRA) that uses probabilistic spin logic devices to realize deep belief network (DBN) architectures while leveraging deterministic computing paradigms to achieve in-circuit training and evaluation. Most of the previous DBN research has focused on software implementations, which provides flexibility, but requires significant execution time and energy due to large matrix multiplications that are relatively inefficient when implemented on standard Von-Neumann architectures. Previous hardware-based implementation of RBM have sought to overcome software limitations by using FPGAs \cite{kim2010large,le2010high}, stochastic CMOS \cite{ardakani2017vlsi}, and hybrid memristor-CMOS designs \cite{bojnordi2016memristive}. Recently, Zand et al. \cite{ZandGLSVLSI} utilized a spintronic device that leverages intrinsic thermal noise within low energy barrier nanomagnets to provide a natural building block for RBMs. While most of the aforementioned designs only focus on the test operation, the work presented herein concentrates on leveraging technology heterogeneity to implement a train and evaluation circuitry for DBNs with various network topologies on our proposed SNRA fabric.  



 

\begin{figure}
\centering
\includegraphics[scale=0.42]{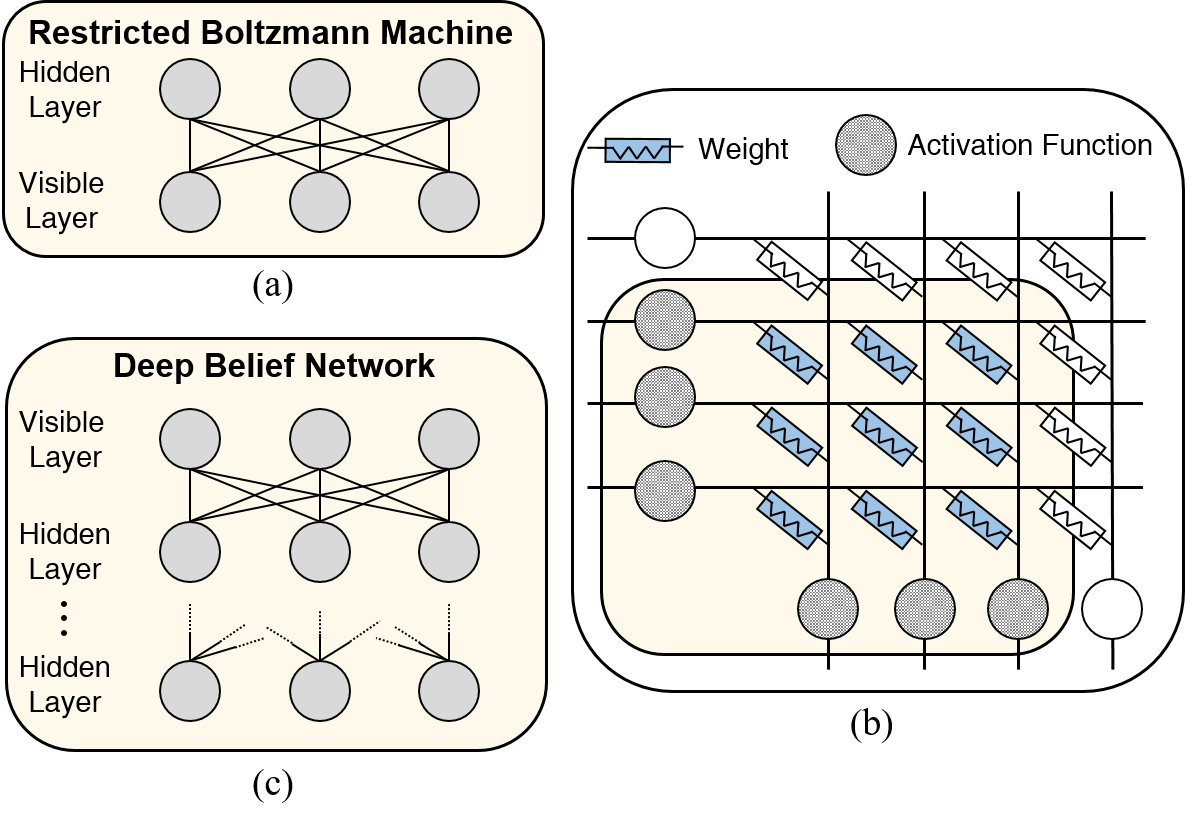}
\caption{(a) An RBM structure, (b) a 3$\times$3 RBM implemented by a 4$\times$4 crossbar architecture, (c) a DBN structure including multiple hidden layers.}
\label{fig:crossbar}
\end{figure}

\section{Restricted Boltzmann Machines}
Restricted Boltzmann machines (RBMs) are a class of recurrent stochastic neural networks, in which each state of the network, \textit{k}, has an energy determined by the connection weights between nodes and the node bias as described by (1), where $s_i^k$ is the state of node \textit{i} in \textit{k}, \textit{b\textsubscript{i}} is the bias, or intrinsic excitability of node \textit{i}, and \textit{w\textsubscript{ij}} is the connection weight between nodes \textit{i} and \textit{j} \cite{ackley1985}.
\begin{equation}
  E(k) = -\sum_{i} s_i^k b_i -\sum_{i<j} s_i^k s_j^k w_{ij} 
\end{equation}

Each node in a RBM has a probability to be in state one according to (2), where $\sigma$ is the sigmoid function. RBMs, when given sufficient time, reach a Boltzmann distribution where the probability of the system being in state \textit{\textbf{v}} is found by (3), where \textit{\textbf{u}} could be any possible state of the system. Thus, the system is most likely to be found in states that have the lowest associated energy.

\begin{equation}
  P(s_i = 1) = \sigma (b_i + \sum_{j} w_{ij} s_j)
\end{equation}

\begin{equation}
	P(v) = \frac{e^{-E(v)}}{\sum_{u} e^{-E(u)}}
\end{equation}

Restricted Boltzmann machines (RBMs) are constrained to two fully-connected non-recurrent layers called the \textit{visible layer} and the \textit{hidden layer}. RBMs can be readily implemented by a crossbar architecture, as shown in Fig.\ref{fig:crossbar}. The most well-known approach for training RBMs is contrastive divergence (CD), which is an approximate gradient descent procedure using Gibbs sampling \cite{carreira2005}. CD operates in four steps as described below:

1. \textit{Feed-forward:} the training input vector, \textbf{$v$}, is applied to the visible layer, and the hidden layer, \textbf{$h$}, is sampled. 

2. \textit{Feed-back:} The sampled hidden layer output is fed-back and the generated input is sampled, \textbf{$v'$}. 

3. \textit{Reconstruct:} \textbf{$v'$} is applied to the visible layer and the reconstructed hidden layer is sampled to obtain \textbf{$h'$}. 

4. \textit{Update:} The weights are updated according to (4), where $\eta$ is the learning rate and \textbf{$W$} is the weight matrix.

\begin{equation}
	\Delta W = \eta (vh^T-v'h'^T)
\end{equation}

RBMs can be readily stacked to realize a DBN, which can be trained similar to RBMs. Training a DBN involves performing CD on the visible layer and the first hidden layer for as many steps as desired, then fixing those weights and moving up a hierarchy as follows. The first hidden layer is now viewed as a visible layer, while the second hidden layer acts as a hidden layer with respect to the CD procedure identified above. Next, another set of CD steps are performed, and then the process is repeated for each additional layer of the DBN.

\begin{figure}
\centering
\includegraphics[scale=0.50]{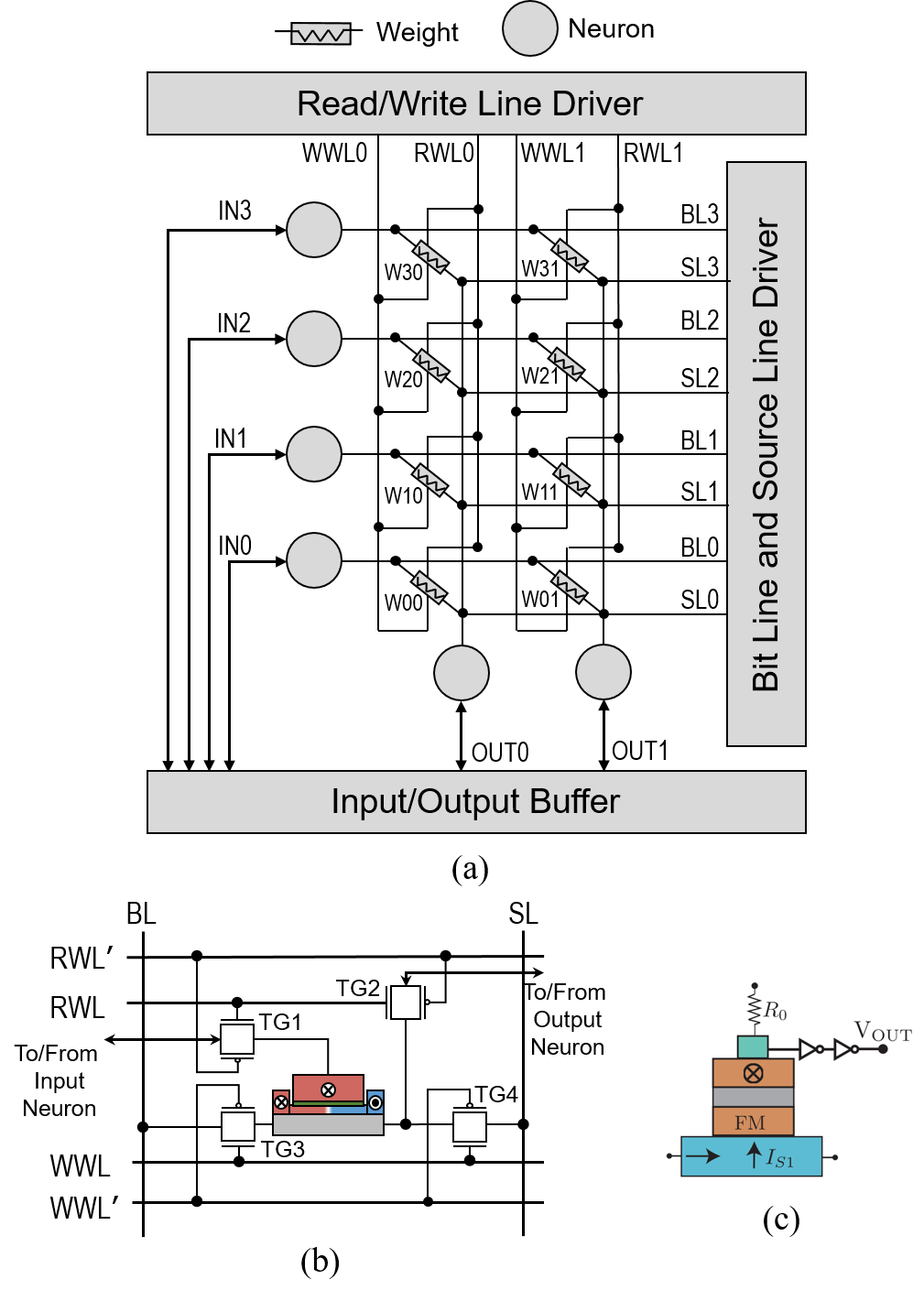}
\caption{(a) A 4$\times$2 RBM hardware implementation, (b) SHE-DWM based weighted connections, and (c) p-bit based probabilistic neuron \cite{Camsari2017}.}
\label{fig:array}
\end{figure}

\section{Proposed RBM Structure}
A feasible hardware implementation of a 4$\times$2 RBM structure is shown in Fig.~\ref{fig:array}(a), in which three terminal spin Hall effect (SHE)-driven domain wall motion (DWM) device \cite{Sengupta2016} is used as weights and biases, while the probabilistic spin logic devices (p-bits) are utilized to produce a probabilistic output voltage that has a sigmoid relation with the input currents of the devices, as shown in Fig.~\ref{fig:array}(b) and Fig.~\ref{fig:array}(c), respectively. The p-bit device consists of a SHE-driven magnetic tunnel junction (MTJ) with a circular near-zero energy barrier nanomagnet, which provides a natural sigmoidal activation function required for DBNs as studied in \cite{Camsari2017,Faria2017,sutton2017,behin2016}. Transmission gates (TGs) are used within the bit cell of the weighted connections to adjust the weights by changing the domain wall (DW) position in SHE-DWM devices, as well as controlling the RBM operation phases. TGs can provide an energy-efficient and symmetric switching behavior \cite{zandTVLSI2017}, which is specifically desired during the training operation.

\begin{table}[]
\centering
\caption{Required signaling to control the RBM operation phases.}
\label{tab:signaling}
\begin{tabular}{llllll}
\hline
\multicolumn{2}{l}{Operation Phase}       & WWL                  & RWL                  & BL                    & SL                    \\ \hline 
\multicolumn{2}{l}{Feed-Forward / Test}          & \multirow{3}{*}{GND} & \multirow{3}{*}{VDD} & \multirow{3}{*}{Hi-Z} & \multirow{3}{*}{Hi-Z} \\
\multicolumn{2}{l}{Reconstruct}           &                      &                      &                       &                       \\
\multicolumn{2}{l}{Feed-Back}             &                      &                      &                       &                       \\ \hline
\multirow{2}{*}{Update} & Increase Weight & \multirow{2}{*}{VDD} & \multirow{2}{*}{GND} & Vtrain  & GND   \\
                        & Decrease Weight &                      &                      & GND                   & Vtrain                   \\   \hline
\end{tabular}
\end{table}

Table~\ref{tab:signaling} lists the required signaling to control the RBM's training and test operations. During the feed-forward, feed-back, and reconstruct operations, write word line (WWL) is connected to ground (GND) and the bit line (BL) and source line (SL) are both in high impedance (Hi-Z) state disconnecting the write path. The read word line (RWL) is connected to VDD, which turns ON the read TGs in the weighted connection bit cell shown in Fig.~\ref{fig:array}(b). The voltage applied by the input neuron generates a current through TG1 and TG2, which is then injected to the output neuron and modulates the output probability of the p-bit device. The amplitude of the current depends on the resistance of the weighted connection which is defined by the position of the DW in the SHE-DWM device. 

During the update phase, the RWL is connected to GND, which turns off TG1 and TG2 and disconnects the read path. Meanwhile, the WWL is set to VDD which activate the write path. Resistance of the weighted connections can be adjusted by the BL and SL signals, as listed in Table~\ref{tab:signaling}. The amplitude of the training voltage (Vtrain) connected to  BL and SL should be designed in a manner such that it can provide the desired learning rate, $\eta$, to the training circuit. For instance, a high amplitude $Vtrain$ results in a significant change in the DW position in each training iteration, which effectively reduces the number of different resistive states that can be realized by the SHE-DWM device.  On the other hand, a higher SHE-DWM resistance leads to a smaller current injected to the p-bit device. Thus, the input signal connected to the weighted connection with higher resistance will have lower impact on the output probability of the p-bit device, representing a lower weight for the corresponding connection between the input and output neurons.

\begin{figure}
\centering
\includegraphics[scale=0.48]{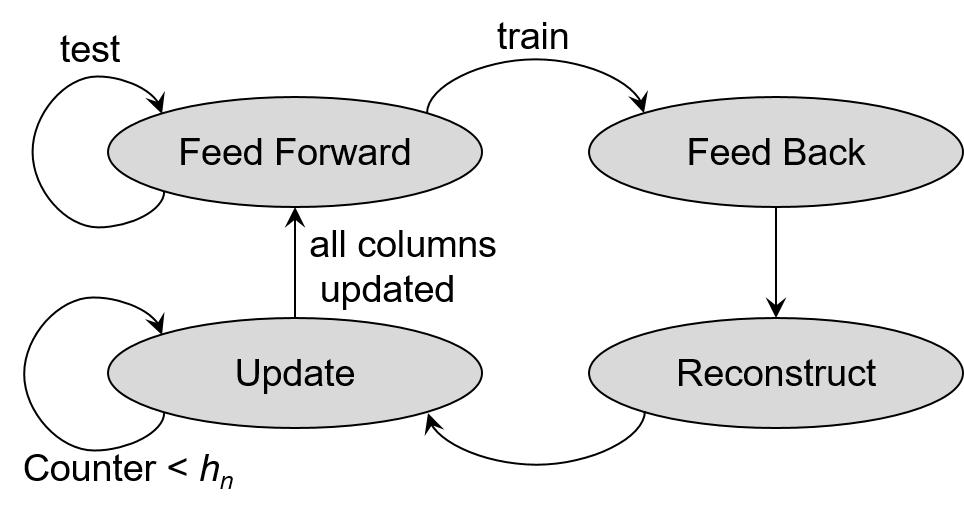}
\caption{FSM designed to control the train and test operations in a DBN.}
\label{fig:FSM}
\end{figure}

\begin{figure*}
\centering
\includegraphics[scale=0.70]{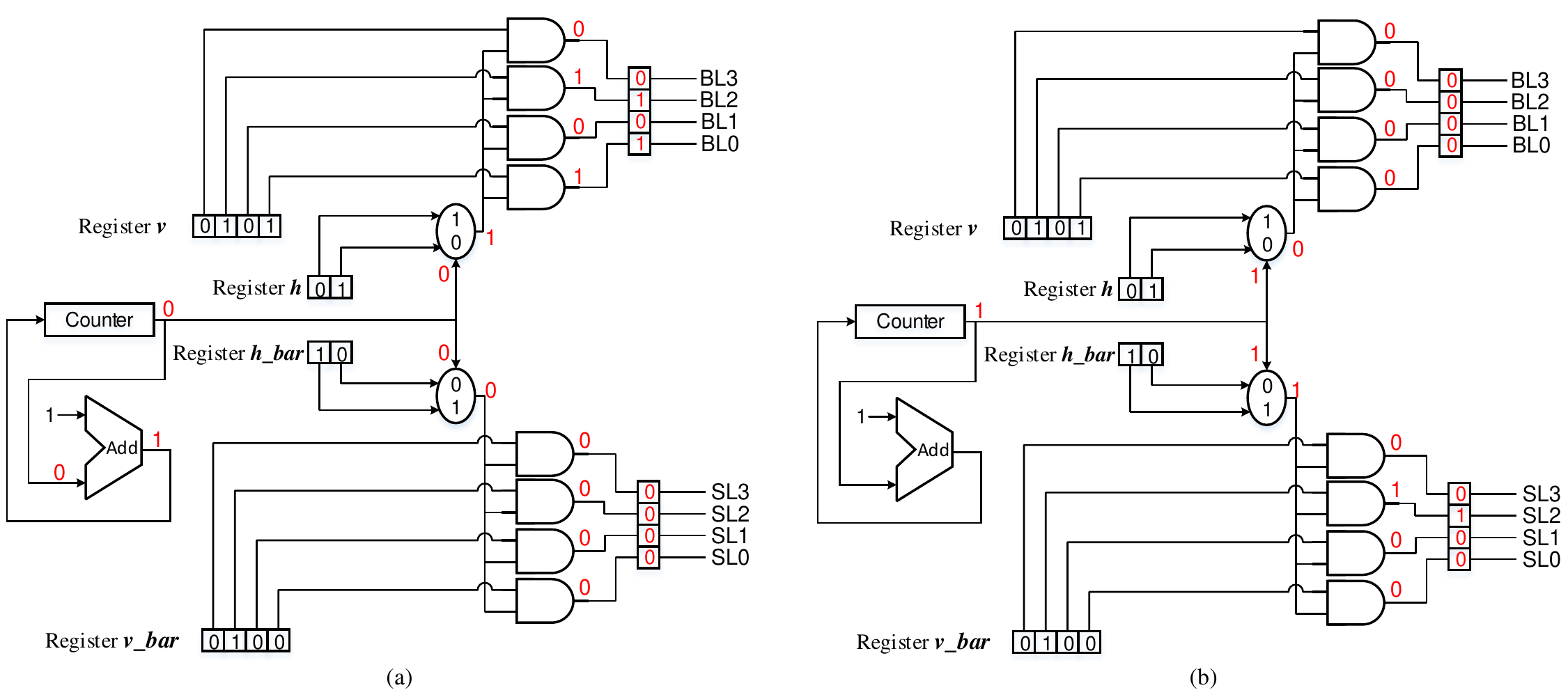}
\vspace{-0.3cm}
\caption{The hardware realization for the \textit{update} state in the FSM developed to train a 4$\times$2 RBM, (a) first clock cycle, and (b) second clock cycle.}
\label{fig:update}

\end{figure*}

\section{Proposed Hardware Implementation of Contrastive Divergence Algorithm}
To implement the contrastive divergence (CD) algorithm required for training the weights in an RBM structure, we have designed a four-state finite state machine (FSM) as shown in Fig.~\ref{fig:FSM}. The proposed FSM is in the \textit{feed-forward} state during the test operation. When the training begins, the input of the visible layer and the corresponding output of the hidden layer will be stored in the \textit{\textbf{v}} and \textit{\textbf{h}} registers, respectively. The size of the \textbf{\textit{v}} and \textbf{\textit{h}} registers depend on the number of neurons in the visible and hidden layers. For instance, in the sample 4$\times$2 RBM shown in Fig.~\ref{fig:array} the size of the \textit{\textbf{v}} and \textbf{\textit{h}} registers are 4-bits and 2-bits, respectively. In the \textit{feed-back} state, the sampled hidden layer is fed-back to the RBM array and the corresponding output of the visible layer is stored in the \textbf{\textit{v\_bar}} register. Next, the stored values in \textbf{\textit{v\_bar}} are applied to the RBM to reconstruct the hidden layer, and the obtained output of the hidden layer will be stored in \textbf{\textit{h\_bar}} register. Finally in the \textit{update} state, the data stored in \textit{\textbf{v}}, \textbf{\textit{h}}, \textbf{\textit{v\_bar}}, and \textbf{\textit{h\_bar}} registers are used to provide the required BL and SL signals to adjust the weights according to (4).

Figure~\ref{fig:update} depicts the schematic of the hardware designed for the \textit{update} state of the FSM developed for a 4$\times$2 RBM. In each clock cycle, The designed circuit adjusts the weights in a single column of the RBM shown in Fig.~\ref{fig:array}. Thus, the number of clock cycles required to complete the update state depends on the number of neurons in the hidden layer of the RBM. A \textit{counter} register is used in the design to ensure that all of the columns in the RBM are updated. The counter value starts from zero and will be incremented in each clock cycle until it reaches the $h_n$ value, which is the total number of nodes in the hidden layer. Once the counter reaches $h_n$, the update state is completed and the FSM goes to the \textit{feed-forward} state. The logical AND gates are used to implement the $vh^T$ and $v'h'^T$ expressions required to find $\Delta W$ for the weights in each column. The output of Boolean gates implementing $vh^T$ and $v'h'^T$ are stored in \textit{\textbf{BL\_reg}} and \textbf{\textit{SL\_reg}} registers, respectively, which provide the required signaling for adjusting the weights according to the Table~\ref{tab:signaling}.

Herein, to better understand the functionality of the hardware developed for the \textit{update} state, we have used an example with the $v$, $h$, $v'$, and $h'$ matrices having the hypothetical values mentioned below:
$$
v=
\begin{bmatrix}
    v_0\\
    v_1\\
    v_2\\
    v_3
\end{bmatrix}
=
\begin{bmatrix}
    1\\
    0\\
    1\\
    0
\end{bmatrix}
\quad
h=
\begin{bmatrix}
    1 \\ 0
\end{bmatrix}
\quad
v'=
\begin{bmatrix}
    0\\
    0\\
    1\\
    0
\end{bmatrix}
\quad
h'=
\begin{bmatrix}
    0 \\  1
\end{bmatrix}
$$
Hence, the $\Delta W$ can be calculated using (4) as shown below:
$$
\Delta W= \eta (vh^T-v'h'^T) = \eta
\begin{bmatrix}
    1 & 0\\
    0 & 0\\
    1 & -1\\
    0 & 0
\end{bmatrix}
=
\begin{bmatrix}
    \delta w_{00} & \delta w_{01}\\
    \delta w_{10} & \delta w_{11}\\
    \delta w_{20} & \delta w_{21}\\
    \delta w_{30} & \delta w_{31}
\end{bmatrix}
$$
\begin{figure}
\centering
\includegraphics[scale=0.6]{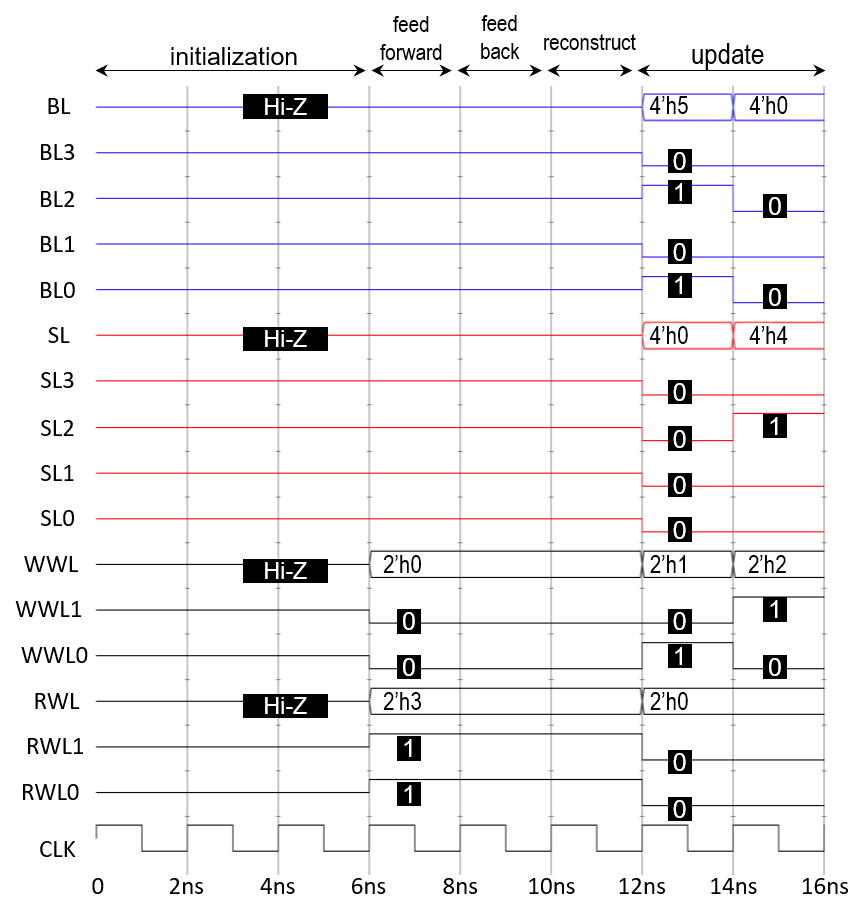}
\caption{The output signals generated by the proposed FSM. The clock frequency is 500MHz, which can be modified based on the design requirements.}
\label{fig:wave}
\end{figure}

\begin{figure*}
\centering
\includegraphics[scale=0.48]{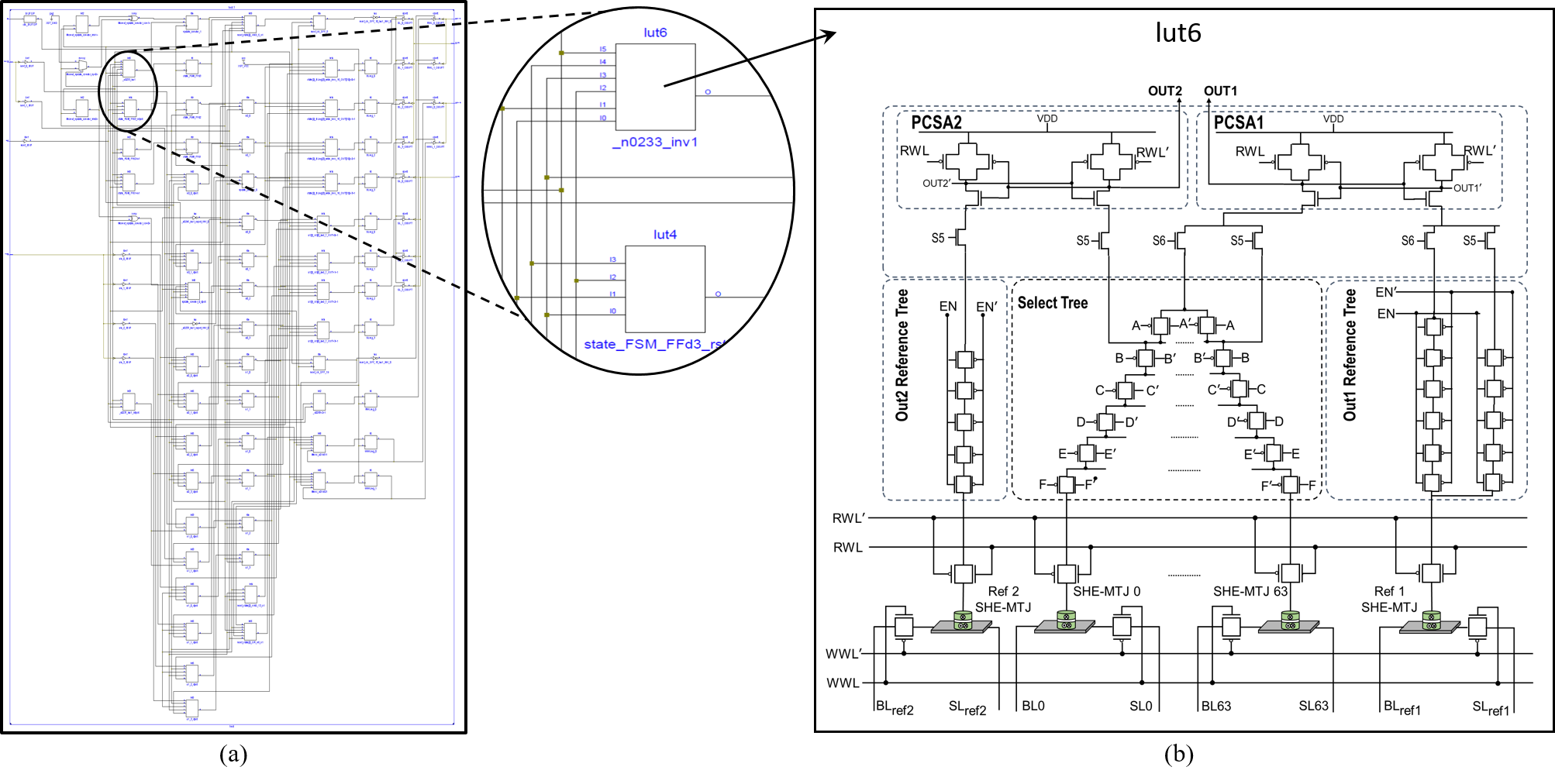}
\caption{(a) The schematic of the hardware designed to control the testing and training operations of a 4$\times$2 RBM implemented on a Xilinx Kintex-7 FPGA family, (b) the structure of a 6-input SHE-MTJ based fracturable LUT used as the building block of the proposed SNRA architecture.}
\label{fig:schematic}
\end{figure*}

According to the obtained $\Delta W$, $w_{21}$ should be decreased while the $w_{00}$ and $w_{20}$ increases, and the remaining weight values remain unchanged. The hardware realization of the mentioned example is shown in Fig.~\ref{fig:update}, in which the values stored in the registers are \textit{\textbf{v}}=4'b0101, \textbf{\textit{h}}=2'b01, \textbf{\textit{v\_bar}}=4'b0100, and \textbf{\textit{h\_bar}}=2'b10. It is worth noting that, the $v_0$ element in the $v$ matrix is stored in the least significant bit of the \textit{\textbf{v}} register, while $v_3$ is stored in the most significant bit. Other matrices are stored to their corresponding registers in the similar manner. In this example, RBM has two output neurons, therefore $h_n$ is equal to two and the update operation can be completed in two clock cycles. In the first cycle shown in Fig.~\ref{fig:update}(a), the counter is equal to zero and the first bits of \textbf{\textit{h}} and \textbf{\textit{h\_bar}} registers are selected by the multiplexers to be used as the input of the AND gates. Therefore, the below BL and SL signals are generated,
$$
BL=
\begin{bmatrix}
    BL0\\
    BL1\\
    BL2\\
    BL3
\end{bmatrix}
=
\begin{bmatrix}
    1\\
    0\\
    1\\
    0
\end{bmatrix}
\quad
SL=
\begin{bmatrix}
    SL0\\
    SL1\\
    SL2\\
    SL3
\end{bmatrix}
=
\begin{bmatrix}
    0\\
    0\\
    0\\
    0
\end{bmatrix}
 $$   
As listed in Table~\ref{tab:signaling}, the above BL and SL signals will increase $w_{00}$ and $w_{20}$ weights shown in Fig.~\ref{fig:array}, if the WWL0 and WWL1 signals are ``1'' and ``0'', respectively. Similarly, in the second clock cycle, the counter is equal to one and the second bits of \textbf{\textit{h}} and \textbf{\textit{h\_bar}} registers are used to produce below BL and SL signals as below, 
$$
BL=
\begin{bmatrix}
    BL0\\
    BL1\\
    BL2\\
    BL3
\end{bmatrix}
=
\begin{bmatrix}
    0\\
    0\\
    0\\
    0
\end{bmatrix}
\quad
SL=
\begin{bmatrix}
    SL0\\
    SL1\\
    SL2\\
    SL3
\end{bmatrix}
=
\begin{bmatrix}
    0\\
    1\\
    0\\
    0
\end{bmatrix}
 $$

This results in a decrease in the $w_{21}$ weight, while the other weights remain unchanged. Thus, the proposed hardware provides the desired functionality required for the \textit{update} state according to (4).  

Herein, we have used the Verilog hardware description language (HDL) to implement our proposed four-state FSM. The ModelSim simulator is used to simulate the developed register-transfer level (RTL) Verilog codes. Figure~\ref{fig:wave} shows the obtained waveforms required for training a 4$\times$2 RBM array with the hypothetical register values mentioned above. The results show that the desired BL, SL, RWL, and WWL control signals are generated in five clock cycles, which verifies the functionality of our proposed FSM.   

To obtain the hardware resources required for our proposed DBN control circuitry, we have synthesized and implemented it using Xilinx ISE Design Suite 14.7. The schematic of the hardware developed to control the testing and training operations for a 4$\times$2 RBM is shown in Fig.~\ref{fig:schematic}(a), in which 32 six-input fracturable look-up table (LUT) and Flip Flop (FF) pairs are used to implement both sequential and combinational logic. It is worth noting that out of the 32 LUT-FF pairs only three of them are utilized for the test operation, thus roughly 90\% of the circuit can be power-gated during the test operation. However in conventional homogeneous technology  FPGAs, volatile static random access memory (SRAM) cells are employed in LUTs to store the logic function configuration data. Therefore, by power-gating the SRAM-based LUTs the configuration data will be lost and the FPGA is required to be re-programmed. In addition to volatility, SRAM cells also suffer from high static power and low logic density \cite{Kuon2007}. Hence, alternative emerging memory technologies have been attracting considerable attention in recent years as an alternative for SRAM cells.

\section{The proposed SNRA architecture}
Herein, we propose a heterogeneous-technology spintronic neuromorphic reconfigurable array (SNRA), which can combine both deterministic and probabilistic logic paradigms. The SNRA fabric is organized into islands of probabilistic modules surrounded by Boolean configurable logic blocks (CLBs). Both the probabilistic and deterministic elements are field programmable using a configuration bit-stream based on conventional FPGA programming paradigms.

Herein, the probabilistic modules consist of RBMs, which can be connected hierarchically within the field-programmable fabric to form various topologies of DBNs. Each RBM leverages SHE-MTJs with unstable nanomagnets ($\Delta \ll 40kT$) to generate the probabilistic sigmoidal activation function of the neurons. With respect to the deterministic logic, the CLBs are comprised of LUTs which realize the training and evaluation circuitry. Non-volatile high energy barrier ($\Delta \geq 40kT$) SHE-MTJ devices are used as an alternative for SRAM cells within LUT circuits. The routing networks include routing tracks, as well as switch and connection blocks similar to that of the conventional FPGAs.  The feasibility of integrating MTJs and CMOS technologies in an FPGA chip has been verified in 2015 by researchers in Tohoku University \cite{Suzuki2015}. They have fabricated a nonvolatile FPGA with 3,000 6-input MTJ-based LUTs under 90nm CMOS and 75nm MTJ technologies. The measurement of fabricated devices under representative applications exhibited significant improvements in terms of power consumption and area. Despite the mentioned improvements, the conventional spin transfer torque (STT)-based MTJ devices suffer from high switching energy and reliability issues. Thus, we propose using SHE-MTJ based LUT circuits with reduced switching energy and increased reliability of tunneling oxide barrier \cite{Manipatruni2014}. Readers are referred to \cite{Fong2016} for additional information regarding the STT-MTJ and SHE-MTJ devices.   
  
Figure~\ref{fig:schematic}(b) shows the structure of a six-input SHE-MTJ based fracturable LUT \cite{zandTNANO}, which can implement a six-input Boolean function or two five-input Boolean functions with common inputs. In general, LUT is a memory with $2^m$ cells in which the truth table of an $m$-input Boolean function is stored. The logic function configuration data is stored in SHE-MTJs in form of different resistive levels determined based on the magnetization configurations of ferromagnetic layer in MTJs, i.e parallel configuration results in a lower resistance standing for logic ``0'' and vice versa. The LUT inputs can be considered as the address according to which corresponding output of the Boolean function will be returned through the select tree. The LUT circuit shown in Fig.~\ref{fig:schematic}(b) includes two pre-charge sense amplifiers (PCSAs) that are used to read the logic state of the SHE-MTJs. The PCSA compares the stored resistive value of the SHE-MTJ cells in the LUT circuit with a reference MTJ cell that its resistance is designed between the low and high resistances of the LUT's SHE-MTJ cells. Therefore, if the resistive value of a SHE-MTJ cell in the LUT circuit is greater than the resistance of the reference cell, the output of the PCSA will be ``1'' and vice versa. The readers are referred to \cite{zandTNANO} for additional information regarding the functionality of a SHE-MTJ based LUT circuit.            

\section{Results and Discussions}
\begin{figure}
\centering
\includegraphics[scale=0.7]{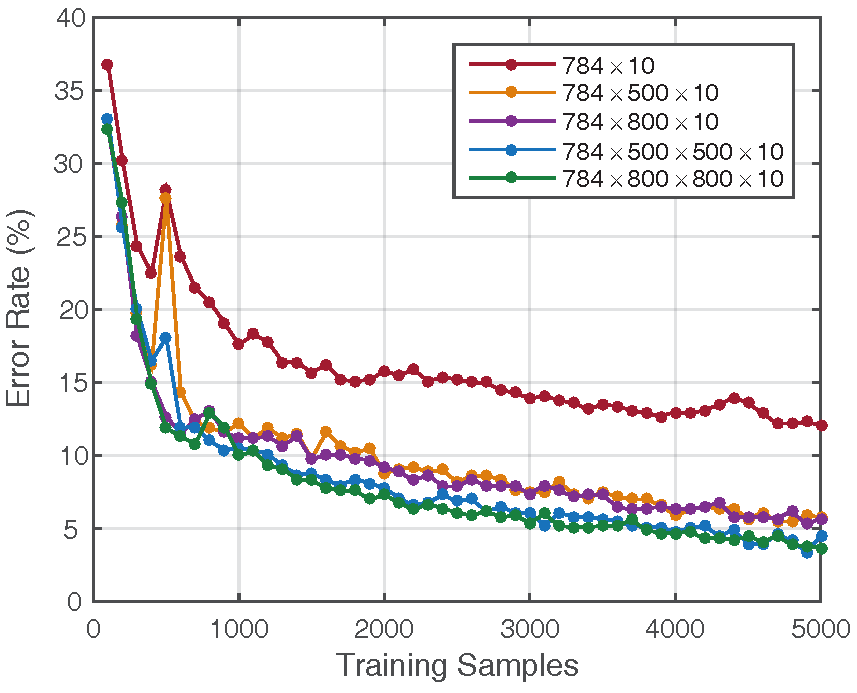}
\caption{Error rate vs. training samples for various DBN topologies \cite{ZandGLSVLSI}.}
\label{fig:err}
\end{figure}

Herein, we have modified a MATLAB implementation of DBN developed in \cite{Tanaka2014} and utilized MNIST data set \cite{Lecun1998} to calculate the error rate and evaluate the performance of our DBN architecture. The simplest model of the belief network that can be used for MNIST digit recognition includes a single RBM with 784 nodes in the visible layer to handle 28×28 pixels of the input images, and 10 nodes in hidden layer representing the output classes. Herein, we have examined the error rate for five different network topologies using 1,000 test samples as shown in Fig.~\ref{fig:err}. As it is expected, increasing the number of the hidden layers, nodes, and training images improves the performance of the DBN, however these improvements are realized at the cost of higher area and power dissipation. 

To compare the resource utilization between the five network topologies investigated in this paper, we have used Xilinx ISE Design Suite 14.7 to implement their control circuitry based on the FSM design proposed in Section IV. The obtained logic resource utilization for each of the mentioned DBN topologies is listed in Table~\ref{tab:resources}. Since the training operation in different layers of the DBN does not happen simultaneously, the resources can be shared for training each RBM. Therefore, the amount of logic resources utilized to implement the FSM of a DBN relies on the size of the largest RBM in the network. For instance, as listed in Table~\ref{tab:resources}, the resource utilization for training a 784$\times$500$\times$10 DBN is equal to that of a 784$\times$500$\times$500$\times$10 DBN, since the size of the largest RBM in both networks is 784$\times$500.          

To provide a fair power consumption comparison between the investigated DBN topologies, we have simulated an SRAM-based six-input fracturable LUT-FF pair in SPICE circuit simulator using 45nm CMOS library with 1V nominal voltage. The obtained static and dynamic power dissipation are listed in Table~\ref{tab:compare}. Herein, we have only focused on the power dissipated by the LUT-FF pairs, and used the below relation to measure the power consumption for each topology:
\begin{equation}
  P_{total} = \sum_{i} A_{i}P_{read} + I_{i}P_{standby}
\end{equation}
where $A_{i}$ and $I_{i}$ are the number of active and idle LUT-FF pairs in RBM $i$ of the DBN, respectively. The obtained power dissipation values for various DBN topologies are listed in the last column of Table~\ref{tab:resources}. The provided trade-offs between the error rate and power consumption can be leveraged to design a desired DBN based on the application requirements.    

\begin{table}[]
\centering
\caption{FSM logic resource utilization and power dissipation for various DBN topologies.}
\label{tab:resources}
\begin{tabular}{lcccc}
\hline
Topology                                                                       & \begin{tabular}[c]{@{}c@{}}Slice \\ Registers\end{tabular} & \begin{tabular}[c]{@{}c@{}}Slice \\ LUTs\end{tabular} & \begin{tabular}[c]{@{}c@{}}Fully-used \\ LUT-FFs\end{tabular} & \begin{tabular}[c]{@{}c@{}}Power \\ Consumption\end{tabular}\\ \hline
784$\times$10                                                                     & 3185            & 123       & 51 & 0.32 mW                                                          \\
784$\times$500$\times$10                                                       & 4655            & 3545       & 1771 & 14.2 mW                                                          \\
784$\times$800$\times$10                                                       & 5533            & 2449       & 2421 & 19.3 mW                                                          \\
784$\times$500$\times$500$\times$10                                            & 4655            & 3545       & 1771 & 25.3 mW                                                          \\
\begin{tabular}[c]{@{}l@{}}784$\times$800$\times$800 $\times$10\end{tabular} & 5617            & 2449       & 2421 & 34.5 mW                                                          \\ \hline
\end{tabular}
\end{table}

\begin{table}[]
\centering
\caption{Performance comparison between six-input fracturable SRAM-based LUT and SHE-MTJ based LUT.}
\label{tab:compare}
\begin{tabular}{llcc}
\hline
\multicolumn{2}{c}{Features}                                                                                        & SRAM-LUT    & SHE-MTJ LUT \\ \hline
\multirow{2}{*}{Device Count}                                                                              & MOS    & 1163              & 565               \\
                                                                                                           & MTJ    & -                 & 66                \\ \hline
\multicolumn{1}{l}{\multirow{3}{*}{\begin{tabular}[l]{@{}c@{}}Power ($\mu$W)\end{tabular}}} & Read   & 6.28              & 1.1               \\
\multicolumn{1}{c}{}                                                                                       & Write  & 28             & 188               \\
\multicolumn{1}{c}{}                                                                                       & Static & 1.6               & 0.21              \\ \hline
\multirow{2}{*}{Delay}                                                                                     & Read   & \textless{} 10 ps  & \textless{} 30 ps  \\
                                                                                                           & Write  & \textless{} 0.1 ns & \textless{} 2 ns   \\ \hline
\multirow{2}{*}{Energy}                                                                                    & Read   & $\sim$ 62.8 aJ     & $\sim$ 33 aJ      \\
                                                                                                           & Write  & $\sim$ 2.8 fJ     & $\sim$ 376 fJ      \\ \hline
\end{tabular}
\end{table}

To investigate the effect of \textit{technology heterogeneity} on the performance of the proposed DBN control circuitry, we have simulated a SHE-MTJ based six-input fracturable LUT in SPICE using 45nm CMOS and 60nm MTJ technologies. The modeling approach proposed in \cite{zandTNANO}\cite{RoohiTCAD} is leveraged to model the behavior of SHE-MTJ devices. In particular, first, a Verilog-A model of the device is developed and used in SPICE to obtain the write current, as well as the power dissipation of the read/write operations. Next, the write current is used in a descriptive MATLAB model of a SHE-MTJ device to extract the corresponding write delay. The simulation results obtained for a SHE-MTJ based six-input fracturable LUT circuit are listed in Table~\ref{tab:compare}.

\begin{figure}[!t]
\centering
\includegraphics[width=3.2in, height=3in]{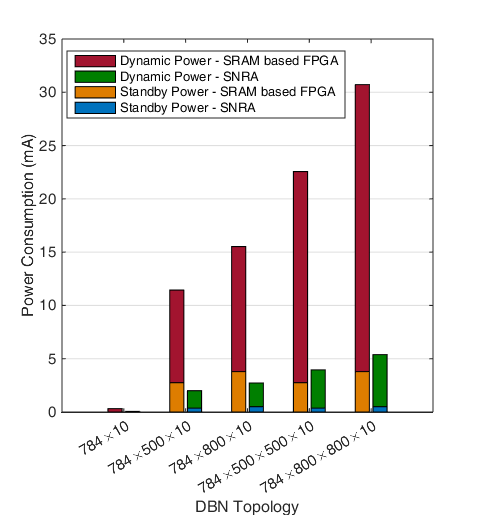}
\caption{Power dissipation of developed FSM for various DBN topologies.}
\label{fig:power}
\end{figure}

Three types of power consumption profiles can be identified in FPGA LUTs. During the configuration phase, the LUTs must be initialized and thus written. This incurs an initial write energy consumption, which occurs infrequently thereafter. Second, upon configuration the LUTs comprising active logic paths will consume read power including a certain sub areas within high gate equivalent capacity of FPGA chips. Third, the remainder of the LUTs, which can be a large number, may be inactive and consume standby power. SRAM-based FPGA is challenged by the difficulty with power-gating LUTs which must retain the stored configuration. While, a SHE-MTJ based LUT can be readily power-gated and incur near-zero standby energy due to its non-volatility characteristic. On the other hand, replacing SRAM cells with SHE-MTJ devices results in a considerable reduction in the transistor count of the LUT circuit since each SRAM cell includes 6 MOS transistors in its structure, while SHE-MTJ devices can be fabricated on top of the MOS circuitry incurring very low area overhead. In particular, SHE-MTJ based LUT circuit achieves at least 51\% reduction in MOS transistor count compared to the conventional SRAM-based LUT, as listed in Table~\ref{tab:compare}. Transistors with minimum feature size are utilized in the SHE-MTJ based LUT circuit to control the SHE-MTJ write and read operations. Thus, the device count results can provide a fair comparison between SHE-MTJ based LUTs and conventional SRAM-based LUTs in terms of area consumption, since all of the MOS transistors used in both designs have the minimum feature size possible by the 45nm CMOS technology.   

Figure~\ref{fig:power} provides a comparison between the conventional SRAM-based FPGA and the proposed SNRA with a focus on the power dissipation induced by LUT-FF pairs utilized to implement the developed DBN control circuitry. The combined improvements in the read and standby modes of the proposed SNRA resulted in realizing at least 80\% reduction in power consumption compared to the conventional CMOS-based reconfigurable fabrics for various DBN topologies. The results obtained for the read operation are comparable to that of the STT-MTJ based FPGA proposed by the Suzuki et al. \cite{Suzuki2015}. However, the utilization of SHE-MTJ based LUTs within the SNRA architecture instead of STT-MTJs can result in at least 20\% reduction in configuration energy as demonstrated by authors in \cite{zandTNANO}.

\section{Conclusion}
The concept of SNRA offers an intriguing architectural approach to realize beyond von-Neumann paradigms which embrace both probabilistic and Boolean computation. As developed herein, the inclusion of in-field programmability offers several practical benefits beyond simulation towards a feasible post-Moore fabric. Most importantly, it can accommodate process variation issues that would otherwise preclude the validity of the baseline training values that differ from the manufactured component. 
 
To coordinate training, a four-state FSM is shown to be sufficient to implement the contrastive divergence (CD) algorithm, as well as the control circuitry for the test operation of DBNs with various topologies. The proposed FSM is capable of unsupervised training of an RBM in $N+3$ clocks where $N$ denoted the number of nodes in the hidden layer of RBM. Interpolating the synthesis results from the Xilinx toolchain indicate a conventional FPGA footprint can accommodate training circuitry for significantly deeper belief networks. This is facilitated using the flexible allocation and routing of layers and their downstream destinations which is a central tenant of CD training. For instance, it was shown that the FSM for both 784$\times$500$\times$10 and 784$\times$500$\times$500$\times$10 DBN topologies can be implemented with 1,771 LUTs, since the size of the largest RBM in both networks is 784$\times$500.           
    
Beyond the flexible architectural approach, within the SNRA fabric, the device parameters are tuned to realize either stochastic switching or deterministic behavior. In particular, near-zero energy barrier SHE-MTJ devices are used to provide a natural probabilistic sigmoidal function required for implementation of the neuron's activation function within an RBM structure. Meanwhile, non-volatile SHE-MTJ devices with high energy barrier ($\Delta \geq 40kT$) can be used to implement LUTs. Use of SHE-MTJ based LUTs achieves more than 80\% and 50\% reduction in terms of power dissipation and area, respectively, compared to conventional SRAM-based reconfigurable fabrics. These improvements are achieved at the cost of higher energy consumption during the reconfiguration operation, which occurs rarely and can be tolerated due to the significant area and power reductions realized during the normal operation of the SNRA.


\section*{Acknowledgment}
This work was supported in part by the Center for Probabilistic Spin Logic for Low-Energy Boolean and Non-Boolean Computing  (CAPSL), one of the Nanoelectronic Computing Research (nCORE) Centers as task 2759.006, a Semiconductor Research Corporation (SRC) program sponsored by the NSF through CCF 1739635.

\bibliographystyle{IEEEtran}
\bibliography{ref}
%



\end{document}